# Surface enhanced covalency and Madelung potentials in Nb doped SrTiO$_3$ (100), (110) and (111) single crystals

G.M. Vanacore[1], L.F. Zagonel[2], N. Barrett[2]

[1] *Dipartimento di Fisica, Politecnico di Milano, Piazza Leonardo da Vinci 32, 20133 Milan, Italy*

[2] *CEA DSM/IRAMIS/SPCSI, CEA Saclay, 91191 Gif sur Yvette, France*

Abstract

The influence of surface enhanced covalency on the Madelung potential is experimentally investigated using angle-resolved photoemission for (100), (110) and (111) SrTiO3 surfaces after annealing in UHV at 630 °C. Deconvolution of the core level spectra (O 1s, Sr 3d and Ti 2p) distinguishes bulk and surface components, which are interpreted in terms of surface enhanced covalency (SEC). By comparing the experimentally measured binding energies with theoretical calculations developed in the framework of the Localized-Hole Point-Ion Model, we quantitatively determine the effective electron occupancy at bulk and surface Sr and Ti sites. Our results confirm the essentially ionic character of Sr–O bond and the partially covalent character of Ti–O bond in bulk STO. The cation Ti and Sr electron occupation is greater for all the three surfaces than in the bulk. Surface covalency shifts the Madelung potential at the surface by ΔEM. ΔEM is a minimum for the (111) surface, and increases through (100), attaining a maximum for (110). The angle-resolved valence band spectra and the work function values also confirm this trend. The results are consistent with d–d charge fluctuations dominating at the surface, whereas metal-ligand charge transfers are more energetically favorable in the bulk.

## 1. Introduction

The Madelung potential reflects the cohesive strength of a crystal and as such is one of the most fundamental properties of solids. Its magnitude is determined by both charge and distance, and in ionic solids it will be influenced by both the effective valency and the crystal structure. Fascinating new material properties like metallicity at the interface of insulating oxides [1-3], have renewed interest in the Madelung potential. Very recently Wadati et al. [4] studied the changes of the Madelung potential in strained La$_{0.6}$Sr$_{0.4}$MnO$_3$ by monitoring the core level binding energy shifts. They measured the Madelung potential as a function of strain, but also showed evidence of the influence of covalency in the Mn–O bonds. In bulk single crystals the strain can be considered negligible which should allow one to identify the contribution of covalency to the Madelung potential.

The electrostatic contribution to the surface energy of a slab cut along a polar direction diverges [5], making such surfaces (or interfaces) inherently unstable. Relaxation, reconstruction and charge redistribution may all take place to compensate the surface dipole. In principle this should also change the Madelung potential at the surface with respect to that in the bulk solid [6]. Observing such changes using photoelectron spectroscopy requires measurement of both surface and bulk core level emissions. In the present article we focus on the influence of the surface enhanced covalency on the Madelung potential for (100), (110) and (111) SrTiO$_3$ (STO) surfaces using angle-resolved photoemission.

A degree of covalency in formally ionic compounds, such as the model perovskite oxide SrTiO$_3$, is well known. Piskunov et al. [7], using *ab initio* calculations determined the valence density charge inside the bulk and found that the Sr–O bond is typically ionic, while the Ti–O bond has a partially covalent character. Bocquet et al. [8] using a cluster model, evaluated the effective occupancy of Ti 3$d$ states for different Ti-oxides in which the Ti formal valence is 4+ ($d^0$). For the STO, they predicted an electron occupancy of 1.1 ($d^{1.1}$), and thus an effective valence of 2.9+. Courths et al. [9], using a point-ion model, estimated the effective charge on the Ti ion to be about 2.5+. *Ab initio* cluster calculations performed by Sousa and Illas [10] confirmed this picture also for titanium dioxide, finding the Ti–O bond to have only 55% ionic character and an effective charge of 2.3+. Moreover, different authors [9,11–13] have suggested that the degree of covalency in Ti–O and Sr–O bonds might change at the surface due to the reduced coordination of each cation.

SrTiO$_3$ (STO) belongs to the family of ABO$_3$ perovskite-type oxides (see Fig. 1). At room temperature the lattice parameter, $a_0$, is 3.91 Å [14]. Each Ti atom shows an octahedral coordination, while each Sr atom has a cuboctahedral coordination. Along the [100] direction the lattice can be built up by alternatively stacking apolar SrO and TiO$_2$ planes, and thus the (100) surface can have either a SrO or a TiO$_2$ terminating plane. The [110] and [111] directions are also comprised of alternate, stoichiometric polar planes, which are SrTiO⁴⁺ and O²⁻ for the [110], and SrO⁴⁻ and Ti⁴⁺ for the [111]. However, given their polar nature these surfaces will relax or reconstruct, modifying the surface electronic structure. Indeed, recent structural studies of the (111) [15] and (110) [16] polar surfaces have demonstrated the presence of complex surface reconstructions, providing that the UHV annealing temperature is sufficiently high to allow atomic diffusion.

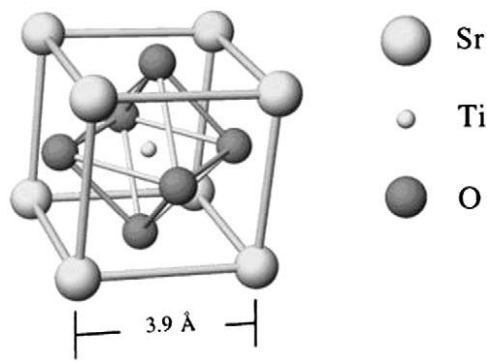

Fig. 1. Unit cell for SrTiO₃ cubic lattice.

In photoelectron spectroscopy the binding energy, $E_B$, of a core electron can be written as:

$$E_B = E_0 + \varepsilon_{sym} - E_M; \qquad (1)$$

in which $E_0$ is the eigenvalue of the spherically symmetric part of the Hamiltonian operator for each ion or free-atom ionization potential, $\varepsilon_{sym}$ is the correction term to the spherical eigenvalues which accounts for the cubic symmetry of the crystal, and $E_M$ is the electrostatic Madelung potential, which describes the inter-atomic interactions. The chemical specificity of each site is included in $E_0$. $\varepsilon_{sym}$ contains the effects of the splitting of $d$ and $p$ orbitals on the core levels. The Madelung potential term in the above equation therefore contains information on the specific crystalline structure of the surface with respect to the bulk and possible changes in bonding covalency at the surface provided of course that the surface core level binding energy can be distinguished from the bulk emission. Knowledge of the chemical potential, which can also be measured using photoemission, will then allow correlation of core level shifts and the Madelung potential.

The appearance of covalency in formally ionic bonding should also influence the electronic structure. It is therefore interesting to relate changes in the Madelung potential with the insulating nature of the transition metal oxide. Many studies of the STO electronic band structure have been carried out, both theoretically [7,17-23] and experimentally [24,25]. The indirect energy band gap is 3.2 eV, and the direct energy band gap is 3.75 eV, as measured by ellipsometry [23]. Zaanen et al. [26] investigated the origin of band gaps and the character of the valence and conduction electron states in 3$d$ transition metal (TM) compounds. They described the physics in terms of: (i) the on-site $d$–$d$ Coulomb repulsion energy $U$, the energy required for charge fluctuations of the type $d_i^n d_j^n \leftrightarrow d_i^{n-1} d_j^{n+1}$; (ii) the charge transfer energy $\Delta$, which describes charge fluctuations of the type $d_i^n \leftrightarrow d_i^{n+1}\underline{L}$, where $\underline{L}$ denotes a hole in anion valence band (ligand p states); and (iii) the ligand p-metal d hybridization energy T.

The late TM compounds mainly fall in the charge transfer regime (Δb U) and the band gap is proportional to $U$ [27], while the early TM compounds (including the STO) were originally classified in the Mott-Hubbard regime ($U$ b Δ) and the band gap is proportional to $U$ [26]. Several authors have reclassified the STO as a charge transfer insulator with a high $p$–$d$ hybridization energy, which will result in strong covalency [8,28,29].

After briefly describing the sample preparation procedure and the experimental details, the core level (O 1$s$, Ti 2$p$ and Sr 3$d$) and valence band photoemission spectra are presented. Deconvolution and fitting of the core level spectra are performed in order to distinguish between bulk and surface components. Comparing the experimen- tally measured binding energies with theoretical calculations devel- oped in the framework of the Localized-Hole Point-Ion Model, we quantify the effective electron occupancy at bulk and surface Sr and Ti sites. The results are confirmed by the valence band spectra and the quantitative determination of the work function for the three surfaces. Finally, we discuss possible charge fluctuations mechanism at the surface with respect to the bulk.

## 2. Experiment

Three commercial SrTiO₃ single crystals (100), (110), and (111), doped with Nb atoms (0.5 wt.%) (SurfaceNet GmbH) were finely ground with diamond paste (smallest grain size: 0.5 μm) and chemo-mechanical polishing, removing 20 μm of the material. Seyton like slurries, containing colloidal SiO₂, KOH or NaOH and some Tensides (pH of the freshly prepared slurry was about 12) were used for the matter. Final surface roughness was 0.2-0.4 nm, as measured by scanning tunnelling microscopy. In order to clean the surface, *in situ* ion etching, oxygen plasma, and annealing have been tested, during which the Sr, Ti, O, and C concentrations were monitored by Auger Electron Spectroscopy (AES). In the case of Ar⁺ ion etching, oxygen was preferentially sputtered leading to the formation of surface vacancies and the increase in surface roughness made carbon removal difficult. Our oxygen plasma efficiently removed the carbon, but altered continuously the Ti and Sr stoichiometries. Annealing at 630 °C was the most effective cleaning procedure, both for carbon removal and for maintaining the oxygen stoichiometry (see Fig. 2a). Preliminary XPS analysis showed no evidence of $Nb_xO_y$ formation after the annealing process excluding the appearance of Nb surface segregation phenomena, and negligible surface carbon contamination as shown in Fig. 2b.

Therefore, annealing at 630 °C for 1 h 30 min in UHV (the pressure was always below 1 · 10⁻⁷ Pa) was adopted before every measure- ment sequence. The moderate temperature minimizes the oxygen vacancy creation and atomic diffusion responsible for previously observed surface reconstructions. Attaining the correct surface stoichiometry whilst minimizing the oxygen vacancies is essential for the comprehension of the intrinsic effects of the surface on the bonding covalency between different atoms. The effectiveness of our treatment is confirmed by the LEED patterns (see Section 3.1), where sharp, bright spots appear, and by UPS Valence Band spectra (see Section 3.3), where the absence of intra-gap surface states indicates a negligible surface carbon contamination and a low concentration of oxygen vacancies [30]. A gold foil was placed in contact with the STO samples in order to provide an energy reference in both XPS and UPS experiments.

Angle-resolved XPS spectra have been acquired using a monochro- matic X-Ray source, and a 125 mm hemispherical analyzer (both Omicron Nanotechnology). The source provides Al $K_\alpha$ (1486.7 eV) radiation with a spot size of 0.75 mm, and a line-width of 250 meV. The electron energy analyzer was operated with an energy pass of 25 eV, entrance slit 1 mm, exit slit 5 mm, and angular acceptance of ±8° (in order to suppress photoelectron diffraction effects). The sample manipulator allowed both polar and azimuthal rotations. XPS spectra of O 1s, Ti 2p and Sr 3d core levels have been acquired for STO (100), (110) and (111), at photoelectron detection angles of θ=0° (normal detection), and θ=60° (grazing detection). The latter gives a twofold increase in surface sensitivity. The bulk Au 4f7/2 peak (at 83.9 eV) was used as the energy reference for the core level spectra. Carbon contamination has been checked by monitoring the C 1s core level peak (at 284.7 eV). A worst case estimation of C atomic concentration on able to rotate around the sample, allowing polar rotation of detection direction without moving the sample. The base pressure in UHV chamber during all measurements was between 1 and 5 × 10$^{-8}$ Pa. UPS spectra have been acquired for (100), (110) and (111) surfaces at detection angles of $\vartheta=0°$ (normal detection) and $\vartheta=66°$ (grazing detection).

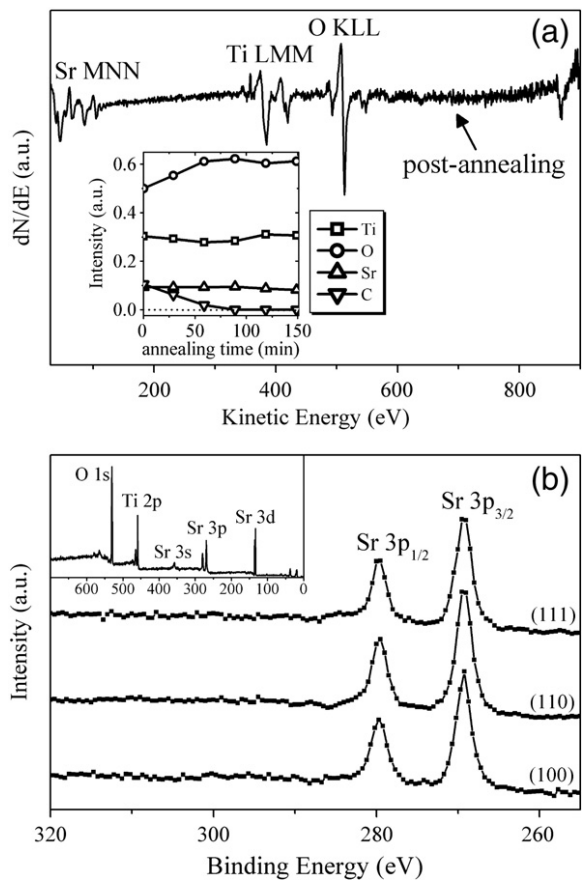

Fig. 2. (a) Main panel: Auger wide spectrum after annealing at 630 °C; inset: Auger intensities Sr, Ti, O, and C as a function of annealing time at 630 °C. (b) XPS Sr 3p spectra and survey spectrum (inset) after annealing at 630 °C showing the absence of carbon contamination.

## 3. Results

### 3.1. Low energy electron diffraction (LEED)

The LEED measurements have been obtained with a standard experimental setup. Fig. 3 shows the raw diffraction patterns (inverted images) for (100), (110) and (111) surfaces, measured at primary beam energy around 100 eV sensitive to the first few atomic layers (the inelastic mean free path (IMFP) of 100 eV electrons in STO is ~ 5 Å [32]). The spots are small and sharp reflecting long range order and high coherence length. The patterns are consistent with a (1 × 1) termination plane for all the three surfaces. Slight charging is observable in the lower part of Fig. 3a for the (100) surface. The absence of higher order reconstructions is consistent with the fact that they are only observed at higher annealing temperature, however we cannot exclude some relaxation particularly for (110) and (111) surfaces [15].

### 3.2. Angle-resolved XPS

Figs. 4, 5 and 6 show the raw data and the corresponding peak deconvolution of the Sr 3d, O 1s and Ti 2p core level spectra for the (100), (110) and (111) surfaces measured at normal and grazing detection. The spectra have been normalized to the maximum intensity. The fitting procedure was performed using a non-linear Gaussian/Lorentzian line shape. A Shirley algorithm [33] has been used for background subtraction in the case of O 1s and Sr 3d spectra, while in the case of Ti 2p an inelastic energy-loss background from experimentally measured electron energy-loss spectra (EELS) (see below) was used to remove extrinsic loss features. Table 1 contains the principal peak parameters of the different components in each spectrum. They are in substantial agreement with data reported in literature for the STO [34-38].

For the Sr 3d spectra, fitting attempts (not shown) using a single doublet yielded unphysical angle-dependent mixing parameters, full width half maximum (FWHM) and branching ratios. A more physically consistent deconvolution of all Sr 3d spectra can be obtained using two overlapping doublets separated by 0.8 ± 0.1 eV for the (100) surface and 0.7 ± 0.1 eV for the (110) and (111) surfaces, with a constant spin-orbit splitting (S.O.S.) equal to 1.8 ± 0.1 eV, a FWHM of 0.9 ± 0.1 eV and a branching ratio of 1.5. The S.O.S value obtained from our fitting is very close to the values reported by the sample surface has been obtained using a discrete-layer model [31], in which the C is assumed to be only at the surface. This gives an atomic carbon concentration between 1–2% of a monolayer after the UHV annealing at 630 °C.

The angle-resolved UPS experiments have been performed using a standard He I discharge lamp (21.2 eV), and a 65 mm hemispherical analyzer (both Omicron Nanotechnology). The photon beam had a spot size of about 2 mm in diameter and a line-width of about 1 meV. The electron analyzer was operated with pass energy of 5 eV, entrance slit 1 mm, and angular acceptance of ±1°, giving an energy resolution of ~ 36 meV. The analyzer is placed inside the UHV chamber, and is several authors [35,36]. The FWHM of each doublet within a given spectrum was the same, since the core hole lifetime should not change significantly between surface and bulk. A further confirmation of the reliability of our deconvolution was the ability to obtain good fits to the spectra at different detection angles simply by changing the relative area between the high binding energy (HBE) doublet (Doublet 1) and the low binding energy (LBE) doublet (Doublet 0). These results suggest that the Sr atoms in STO crystals are in two distinct chemical or structural environments. The relative intensity of the HBE doublet (Doublet 1) increases for the surface sensitive spectrum ($\vartheta$ =60°), whereas that of the LBE doublet (Doublet 0) does not (Table 2). Therefore, the Doublet 0 can be related to the photoelectron signal from bulk Sr atoms, while the Doublet 1 can be related to surface Sr atoms.

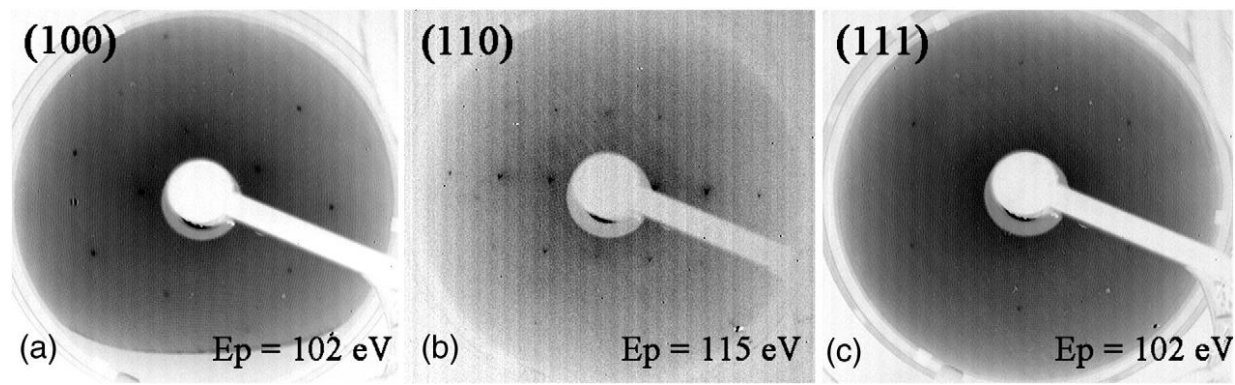

Fig. 3. Experimental LEED patterns (inverted images) measured on: (a) (100) surface, (b) (110) surface, and (c) (111) surface using a primary energy of ~100 eV.

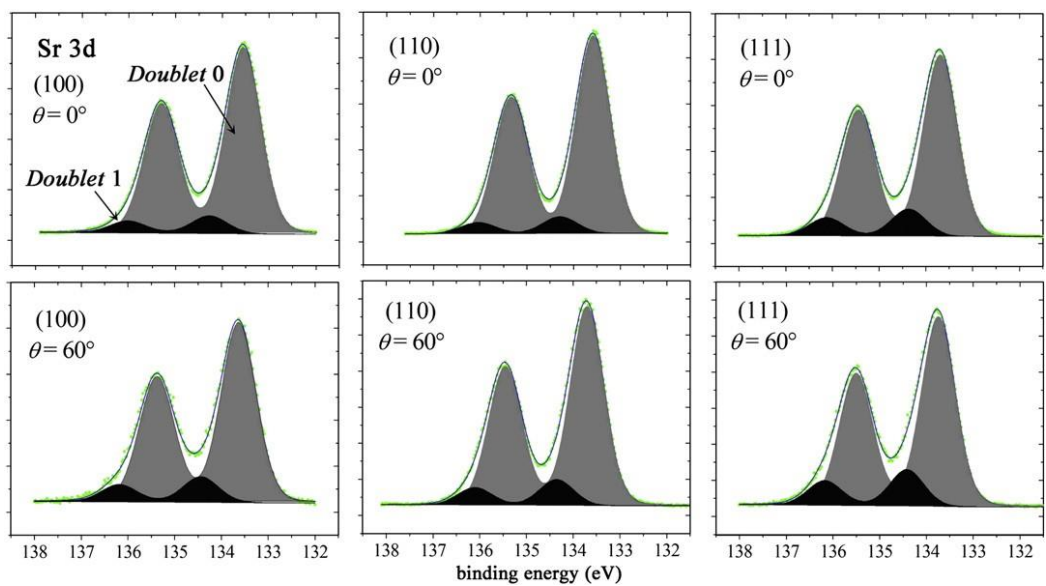

Fig. 4. Sr 3d core level spectra measured for the (100), (110) and (111) surfaces, at normal and grazing detection (60°).

For O 1s, all fitting attempts with less than three peaks led to inconsistent results. Good fits are obtained for the O 1s spectra using three overlapping almost-Gaussian peaks. The HBE peaks, Peak 1 and Peak 2, are shifted with respect to the LBE peak, Peak 0 (at ~530.1 ± 0.1 eV), by 0.7 ± 0.1 eV and 2.1 ± 0.1 eV for the (100), 1.0 ± 0.1 eV and 2.1 ± 0.1 eV for the (110), and 0.9 ± 0.1 eV and 2.1 ± 0.1 eV for the (110) surfaces, respectively. Table 3 shows how the peak intensities change as function of the detection angle: the HBE peaks increase their intensity when the spectrum is acquired in more surface sensitive conditions, while the LBE peak is almost unaffected. Thus, we consider the HBE peaks as a surface signal, while the LBE peak is related to bulk O atoms.

Moreover, the relative intensities of the two surface peaks show no correlation as function of the detection angle, suggesting that both Peak 1 and Peak 2 represent O atoms in distinct surface chemical environments. Emission with a binding energy of Peak 2 (~ 532.1 eV) is frequently reported in literature [36-38] as being characteristic of carbonate compounds or chemisorbed $H_2O$, $OH^-$ or $CO_2$ groups on the sample surface. Due to the low level of carbon contamination on the sample surface (see Section 2) we believe the Peak 2 to originate mainly from chemisorbed $H_2O$ or $OH^-$. On the other hand, the Peak 1, whose intensity variation as function of detection angle is qualitatively quite similar to that of the Sr surface doublet (see Tables 2 and 3), can be associated to O surface atoms in the terminating layer, as recently reported by some of us for polycrystalline STO [38].

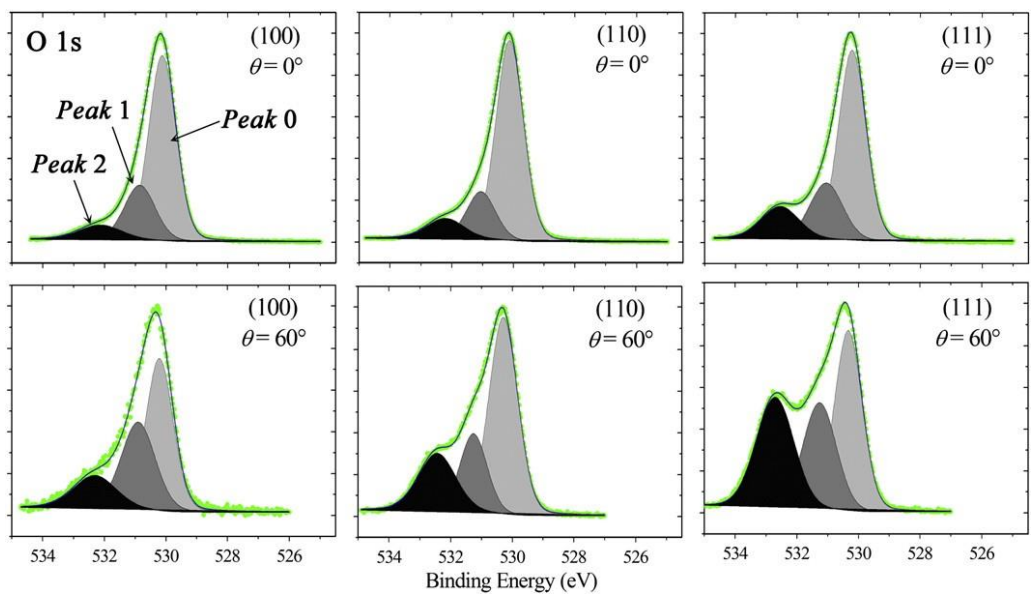

Fig. 5. O 1s core level spectra measured for the (100), (110) and (111) surfaces, at normal and grazing detection (60°).

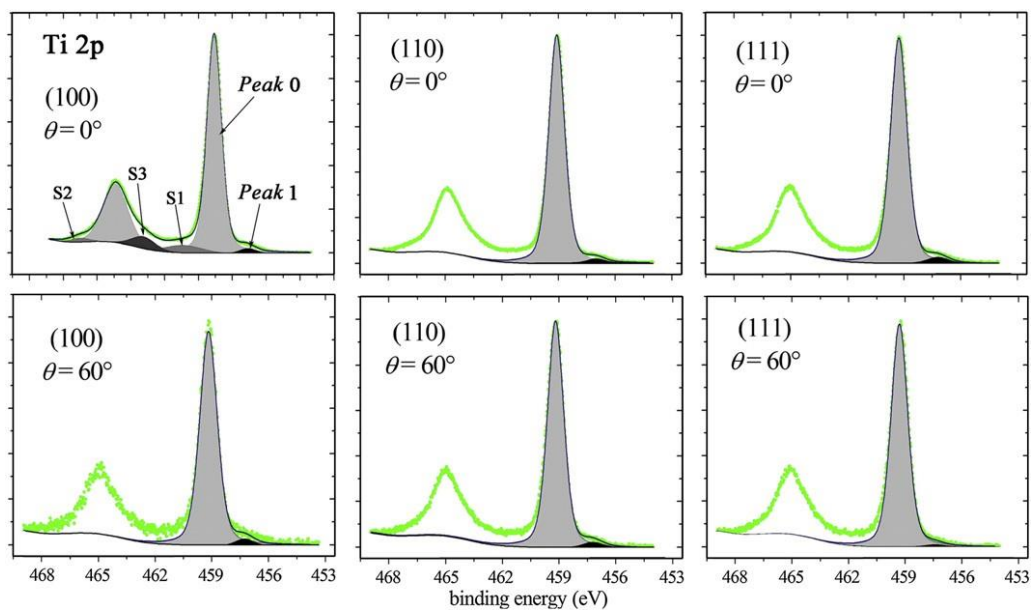

Fig. 6. Ti 2p core level spectra measured for the (100), (110) and (111) surfaces, at normal and grazing detection (60°).

For Ti 2p spectra, the fitting can only be correctly performed after the removal of the inelastic scattering part present in the XPS spectrum. We have measured EELS spectra for the three surfaces (see Fig. 7), and extracted the inelastic energy-loss backgrounds for the experimental photoemission spectra. The EELS spectra were measured at electron kinetic energies matching those of the main photoemission peak (~ 1024 eV), with an analyser band pass set to 10 eV in order to obtain a FWHM of the elastic peak of about 1 eV, matching the FWHM of the $2p_{3/2}$ component in the XPS spectra.

The STO(100) EELS spectrum compares well with data in the literature [23,39]. Below 10 eV the main loss structure is attributed to interband transitions between the O 2p valence electrons to Ti 3d conduction band states. The intensity centred on ~ 14 eV is associated with O 2p resonances [39], whereas Van Benthem et al. [23] identify O 2p →Sr 4d $t_{2g}$ and $e_g$ transitions. The broad loss peak at 30 eV is interpreted as being made up of interband transitions involving O 2s→Ti 3d and Sr 4d levels and the bulk plasmon, whose estimated position from the electron density is ~ 27 eV. The corresponding surface plasmon ($\omega_S = \omega_B / \sqrt{2}$) may well be the resonance around 20 eV, which is more intense for the (111) surface. The no-loss photoemission spectra were obtained by subtracting a loss background obtained from the EELS data at each point in the XPS spectrum proportional to the ratio between the EELS elastic peak and the photoemission intensity (following a procedure similar to that used in [8]). Thanks to this procedure we were able to remove from the XPS spectra the loss feature at ~6.5 eV from the main peak.

Ti 2p spectra can be well represented by a dominant doublet at 459.1 ± 0.1 eV with S.O.S. of 5.8 eV, and by a low intensity structure on low binding energy (LBE) side of the dominant doublet at ~ −2.0 eV for the (100) and (111) surfaces, and −1.9 eV for the (110). The dominant doublet line shape and S.O.S value obtained from the fitting are closely similar to those reported in the literature [8,39]. The dominant doublet is usually associated to Ti ions with a formal valence 4+ (bulk STO), while the structure on LBE side is believed to be the $2p_{3/2}$ peak of the doublet associated to Ti ions with a reduced charge state [40]. This last peak may therefore have two origins: emission from reduced charge state Ti at the surface, and, based on electronegativity arguments, emission from Ti ions oxygen bridged to Nb dopants in the bulk. Table 4 shows the intensity variations as function of the detection angle for the $2p_{3/2}$ peak of the dominant doublet (Peak 0) and for the shoulder at LBE side (Peak 1). Looking closely, Peak 1 is enhanced for surface sensitive grazing angle detection for both (100) and (110) surfaces, whereas it is attenuated for the (111) surface. Thus, for (100) and (110) surfaces Peak 1 could simply be considered to have a surface origin, associated with surface Ti ions in a reduced charge state. The (111) surface also has a significant Peak 1, however, its behaviour as a function of the detection angle and the Sr 3d spectra suggest a bulk origin, and a Sr rather than Ti rich surface termination. Given the doping level, 1% of Ti sites should contain substitutional Nb. The intensity ratio of the two components at normal detection is about 0.03.

A complete fit to the Ti 2p spectrum is shown in Fig. 6 for the (100) surface at normal detection. The FWHM for the spin-orbit components of the dominant doublet are respectively ~ 1 eV and ~ 1.75 eV, with a branching ratio of 2.13. The difference in the FWHM could be due to Coster-Kronig decays of the core hole in the $2p_{1/2}$ level, as reported by Zaanen and Sawatzky [41]. Three additional peaks are present in Ti 2p spectrum at 2.0 ± 0.1 eV (S1), 4.2 ± 0.1 eV (S3) and 7.9 ± 0.1 eV (S2) on the high binding energy side of the $2p_{3/2}$ peak (Peak 0), representing satellite structures due to intrinsic loss processes. The peaks S1 and S2 are similar to those reported by Oku et al. [42], while the peak S3 is reported here for the first time. The physical origin of the latter is unknown and still being studied.

Table 1
Binding energies of the bulk components and surface core level shifts ($\Delta E_B = E_B^{surf} - E_B^{bulk}$) for O 1s, Sr 3d and Ti 2p. The uncertainty is shown in brackets. All values are expressed in eV. For the case of Ti $2p_{3/2}$ at (111) surface (indicated by the symbol *) $\Delta E_B$ does not represent a surface shift, but is due to Ti–O–Nb coordination in the bulk (see text).

|  | Bulk | (100) | (110) | (111) |
|---|---|---|---|---|
|  | $E_B$ | $\Delta E_B$ | $\Delta E_B$ | $\Delta E_B$ |
| O 1s | 530 (0.1) | +0.7 (0.1) | +1.0 (0.1) | +0.9 (0.1) |
|  |  | +2.1 (0.1) | +2.1 (0.1) | +2.1 (0.1) |
| Sr $3d_{5/2}$ | 133.5 (0.1) | +0.8 (0.1) | +0.7 (0.1) | +0.7 (0.1) |
| Ti $2p_{3/2}$ | 459.1 (0.1) | −2.0 (0.1) | −1.9 (0.1) | −2.0 (0.1)* |

Table 2
Intensity of peak components in Sr 3d spectra as a function of the detection angle.

|  | $\dfrac{AreaDouble\ 0\ (\theta = 60°)}{AreaDouble\ 0\ (\theta = 0°)}$ | $\dfrac{AreaDouble\ 1\ (\theta = 60°)}{AreaDouble\ 1\ (\theta = 0°)}$ |
|---|---|---|
| Sr 3d STO(100) | 1,02 | 1,55 |
| Sr 3d STO(110) | 0,9 | 1,4 |
| Sr 3d STO(111) | 0,93 | 1,29 |

Table 3
Intensity of peak components in O 1s spectra as a function of the detection angle.

| | $\frac{AreaPeak\ 0\ (\vartheta = 60°)}{AreaPeak\ 0\ (\vartheta = 0°)}$ | $\frac{AreaPeak\ 1\ (\vartheta = 60°)}{AreaPeak\ 1\ (\vartheta = 0°)}$ | $\frac{AreaPeak\ 2\ (\vartheta = 60°)}{AreaPeak\ 2\ (\vartheta = 0°)}$ |
|---|---|---|---|
| O 1s STO(100) | 0,82 | 1,74 | 2,22 |
| O 1s STO(110) | 1,02 | 1,6 | 2,9 |
| O 1s STO(111) | 0,95 | 1,86 | 3,2 |

Table 4
Intensity of peak components in Ti 2p spectra as a function of the detection angle.

| | $\frac{AreaPeak\ 0\ (\vartheta = 60°)}{AreaPeak\ 0\ (\vartheta = 0°)}$ | $\frac{AreaPeak\ 1\ (\vartheta = 60°)}{AreaPeak\ 1\ (\vartheta = 0°)}$ |
|---|---|---|
| Ti 2p STO(100) | 1.02 | 1.28 |
| Ti 2p STO(110) | 1.00 | 1.44 |
| Ti 2p STO(111) | 0.99 | 0.27 |

For the case of (100) surface the proportion of SrO-terminated and TiO$_2$-terminated regions can be estimated from the relative surface/bulk core level intensities:

$$\%\ SrO = \frac{I_{Sr}^{surf} = I_{Sr}^{STD}}{I_{Sr}^{surf} = I_{Sr}^{STD} + I_{Ti}^{surf} = I_{Ti}^{STD}} \quad \%\ TiO_2 = \frac{I_{Ti}^{surf} = I_{Ti}^{STD}}{I_{Sr}^{surf} = I_{Sr}^{STD} + I_{Ti}^{surf} = I_{Ti}^{STD}}:$$

The standard values, $I^{STD}$, for Ti and Sr are approximated by the bulk intensities: $I_{Ti}^{STD} \approx I_{Ti}^{bulk}$, and $I_{Sr}^{STD} \approx I_{Sr}^{bulk}$. The surface intensities have been corrected in order to take in account the contribution of the bulk emission coming from Ti ions bridged with Nb dopants. This yields 88% SrO-termination and 12% TiO$_2$-termination, in good agreement with a recent photoelectron diffraction study [43], where a comparison between experimental XPD patterns and multiple scattering calculations gave 90% of SrO-termination and 10% of TiO$_2$-termination.

### 3.3. Angle-resolved UPS

Fig. 8a shows the raw data of the normally detected spectra, after background subtraction [44]. The principal structures appearing in the spectra originate from the valence band (VB) region between 3 eV and 9 eV, which corresponds primarily to O 2p non-bonding and bonding states [22]. The experimental VB at normal emission for STO(100) is compared with the Density of States (DOS) calculated by Chambers et al. [45] using a self-consistent GW approach (dotted curve), and the same DOS broadened by convolution with Gaussians of 70 meV (thin solid line) and 1 eV FWHM (thick solid line). The valence band onsets are obtained from a linear extrapolation of the O 2p leading edges (see Fig. 8b). They are practically constant for the three samples, suggesting that the Fermi level is pinned by the bulk Nb doping, just below the conduction band. Thus, the chemical potential μ for all three samples is the same. This will be important in correlating the core level shifts with variations in the Madelung potential. The valence band widths, W, obtained from linear fitting the upper and lower valence band edges (see inset in Fig. 8a) were 5.60, 5.50 and 5.67 eV for the (100), (110) and (111) surfaces, respectively (see Table 5). These values are reasonable since one would expect a greater overlap of the O 2p orbitals for denser surfaces, leading to broadening of the valence band. The valence band widths obtained from the surface sensitive UPS spectra measured along the Sr-O bond direction are 5.52, 5.42 and 5.54 eV for the (100), (110) and (111) faces, while those of the surface sensitive spectra measured along the Ti–O bond direction are 5.64 and 5.70 for the (100) and (110) faces (see Table 5).

The valence band width for STO(100) agrees well with previous results [44]. Along the Sr-O bond direction it is reasonable that a lower coordination reduces the hybridization, giving a narrower valence band. The increase of the bandwidth along the Ti–O bond direction could be due to an enhancement of the p-d hybridization between cations and anions, and thus higher covalency at surface. Again, this suggests a more complex behaviour than simple band narrowing due to overall reduced surface coordination and is developed in the discussion section. The energy band gap region does not show significant surface states (see Fig. 8b), confirming a low

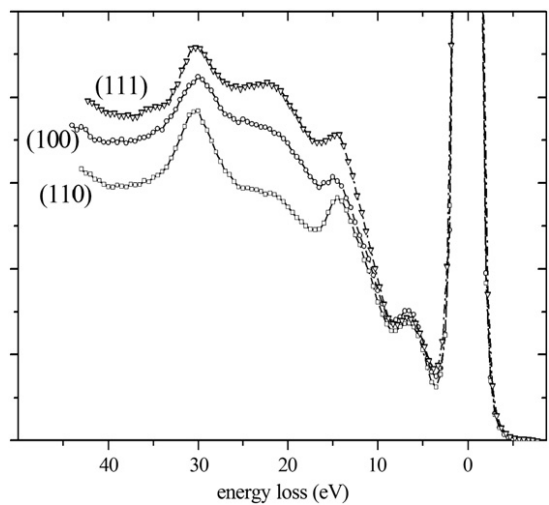

Fig. 7. Electron energy-loss spectra measured at 1024 eV on the STO(100), (110) and (111).

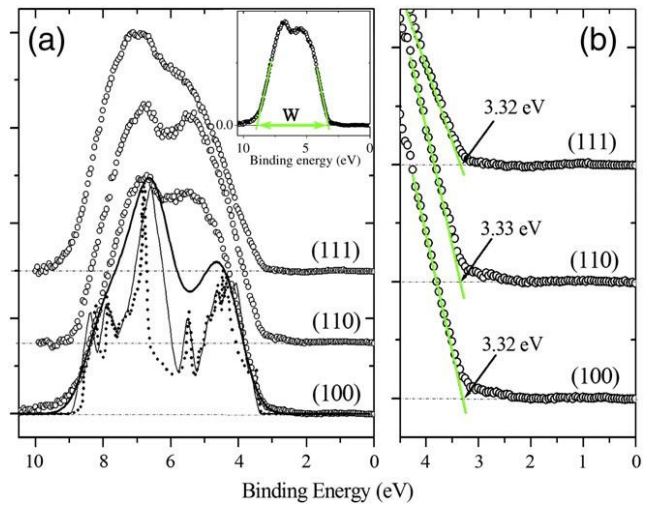

Fig. 8. (a) UPS spectra measured at normal detection for the three surfaces; the experimental VB for STO(100) is compared with the raw (dotted curve) and the properly broadened with FWHM of 70 meV (thin solid line) and 1 eV (thick solid line) scGW DOS [45]; the inset shows the linear fitting of the band edges for the determination of the VB width. (b) Close ups of the valence band maximum (VBM) region for the spectra in (a), indicating the values of VBM as determined by a linear fitting of the upper edge.

Table 5
STO work function, valence band width and valence band maximum for (100), (110) and (111) surfaces.

| Work function (eV) | VB width (eV) | | | VB maximum (eV) | | |
|---|---|---|---|---|---|---|
| | $\theta=0°$ | $\theta=66°$ (Sr-O) | $\theta=66°$ (Ti-O) | $\theta=0$ | $\theta=66°$ (Sr-O) | $\theta=66°$ (Ti-O) |
| (100) 4.49 | 5.60 | 5.52 | 5.64 | 3.33 | 3.45 | 3.35 |
| (110) 4.55 | 5.50 | 5.42 | 5.7 | 3.33 | 3.55 | 3.41 |
| (111) 4.46 | 5.67 | 5.54 | – | 3.32 | 3.61 | – |

concentration of oxygen vacancies [30,46] induced by our annealing process and low surface carbon.

The sample polarization with a potential of −6 V allows to clearly distinguish the photoemission cut-off, determined using a complementary error function fit to the threshold region. In this way, we could evaluate the spectrum width, $\Delta E$, defined as the difference between the Fermi Level and the photoemission cut-off, and by using the formula, $\Delta E=\hbar\omega-\Phi$, the work function, $\Phi$, for each surface (see Table 5). The value of $\Phi$ varies with the crystallographic orientation of the surface ($\Phi^{(111)} < \Phi^{(100)} < \Phi^{(110)}$), known as work function anisotropy [47]. In a recent theoretical study of the work function anisotropy of an oxide surface ($CrO_2$), Attema et al. [48] have suggested a model based on surface stability/valency, atom density and the electronegativity of the surface atoms. The surfaces with the most electropositive element should have the lowest work function. For the (111) surface the lower work function points to a Sr-O rich surface termination. Finally, the pinning of the Fermi level is just below the STO conduction band providing further evidence that our STO surfaces are free from significant band bending, like the fractured surfaces studied by Kohiki et al. [39].

## 4. Discussion

### 4.1. Origin of the surface core level shifts

From LEED patterns, surface relaxation is small and there is no significant long range reconstruction. However, redistribution of electronic charge is indeed likely, particularly for the polar directions. The surface core level shift of titanium, strontium and oxygen are therefore interpreted with the help of a model of surface enhanced covalency (SEC) [11,49]. Within this framework, the charge distribution will change the Madelung potential. In the absence of changes in the chemical potential, the surface core level shifts can be directly related to the value of the Madelung potential at the surface. In the surface plane each cation (Sr and Ti) is coordinated with a lower number of nearest-neighbours anions (O) with respect to the bulk. Since oxygen is more electronegative than strontium and titanium, this induces an increase of the valence electronic charge on each cation site. The effective charge state is thus reduced for both cations and anions, while the bonding charge and the ligand-metal hybridization energy ($T$) become larger. In this sense, surface Sr–O and Ti–O bonds have a more covalent character than in the bulk. The reduction of the effective charge state of the surface ions induces two competitive effects on the core level binding energy positions: (i) on one hand, the modification of the screening reduces the nuclear Coulomb attraction experienced by the core electrons in the case of Sr and Ti, while it is enhanced for O ions; and (ii) on the other hand, the crystalline potential generated by each ion is effectively reduced, and thus the core electrons are more strongly attracted by their own nucleus. Depending on the relative strength of these effects, positive (increase of binding energy) or negative (decrease of binding energy) surface shifts can appear for Sr and Ti, whereas only a positive shift is expected for the O core levels, as observed (see Table 1).

### 4.2. Binding energy calculations

Here we present the basic concepts underlying the O, Ti and Sr core level binding energy calculations. We denote $n$ the electron occupancy of Ti 3d states, $m$ the electron occupancy of Sr 5s states, and with $6-k$ the electron occupancy of O 2p states, which is obviously anti-correlated with those of Sr and Ti. The subscripts $b$ and $s$ indicate bulk and surface sites. First, we calculate the binding energies for Ti $2p_{3/2}$ and Sr $3d_{5/2}$ core levels as a function of the valence electron occupancy for bulk and surface sites. Comparison between the calculated and the experimentally measured binding energies allows deduction of $n_b$, $m_b$, $n_s$ and $m_s$.

The calculations have been performed in the framework of the Localized-Hole Point-Ion (LHPI) model [50], modified to account for the symmetry of the specific site, where the binding energy of a core electron is given by the Eq. (1) (see Section 1). The free-atom ionization potential, $E_0$, and the correction term for cubic symmetry, $\varepsilon_{sym}$, have been calculated using Cowan's program [51], and the specific coordination symmetry is included following Butler's point group notation [52]. $O_h$ symmetry is considered for Ti and Sr bulk sites, with a crystal field splitting for Ti 3d states 10 Dq set to 2.0 eV [53]. For the case of Sr, the crystal field splitting is considered to be about 15% less than for Ti sites [54], and relativistic corrections to the binding energies have been included in the calculations. A first approximation to cation screening is included through a one electron transfer from the ligands.

The Madelung potential, $E_M$, has been calculated using Tosi's method [55]. Charge neutrality for the unit cell in the bulk leads to the following expression for the bulk potential:

$$E_M^{bulk}(n_b, i) = E_M^{bulk}(0, i) \cdot \left(1 - \frac{n_b}{4}\right) \quad (2)$$

in which $E_M^{bulk}(0, i)$ is the bulk Madelung potential at sites $i$ (Ti, Sr) with all ions in their ideal oxidation states ($n_b=0$ and $m_b=n_b/2=0$).

To first order, the bulk-surface Madelung shift can be derived as:

$$\Delta E_M = E_M^{surf}(n_s, i) - E_M^{bulk}(n_b, i) = \Delta E_M(0, i) \cdot \left(1 - \frac{n_b}{4}\right) + C_M \cdot (n_s - n_b) \quad (3)$$

in which $\Delta E_M(0, i) = \frac{1}{2}\left(E_M^{plane}(0, i) - E_M^{bulk}(0, i)\right)$ is the bulk-surface shift of the Madelung potential when $n_b=n_s=0$, $E_M^{surf}(n_s, i)$ is the Madelung potential at the surface of the solid, and $E_M^{plane}(0, i)$ is the contribution to the Madelung energy of the 2D surface plane in the ideal oxidation state. The LHPI model allows explicit inclusion of bulk and surface covalency. The first term in Eq. (3) is zero-order and represents the variation of the Madelung potential due to the different surface geometry and the reduced coordination in the terminating plane with respect to the bulk (this is the equivalent of the term $\Delta V_M$ in eqn. 1 of Wadati et al. [4]); the second term is first order and describes how the Madelung potential can vary as a function of the effective charge state of the surface ions, with $C_M = \frac{\partial E_M^{surf}}{\partial n_s} = -\frac{1}{4}E_M^{plane}(0, i)$. The surface covalency enters into the model through the effective electronic occupation at the surface, $n_s$, thus modifying the Madelung potential.

The symmetry of the crystal field at lattice sites at and near the surface is lower than in the interior of the crystal, and this leads to additional splitting of $d$ and $p$ orbitals. However, we found that the bulk-surface variation of the binding energies induced by this term is always much less than the variation induced by the Madelung energy, and thus in all the calculations we always used the correction term $\varepsilon_{sym}$ obtained for a bulk cubic symmetry. In the single crystals there is no strain contribution to the Madelung potential. The UPS measurements strongly suggest a constant chemical potential, thus $\Delta\mu$ in Wadati et al. [4] can be set to zero. Like these authors, we also neglect the contribution of the extra-atomic screening. This is probably reasonable for the purposes of a comparative study of the surface core level shifts of the three crystal orientations.

*4.3. Bulk and surface Ti and Sr occupancies*

Fig. 9a shows the binding energy of Ti $2p_{3/2}$ calculated for a bulk site as a function of the valence electron occupancy, $n$. Thus, $4 - n$ is the effective valence charge ground state. By comparing these values with the experimentally measured $2p_{3/2}$ binding energy (459.1 eV), $n_b$ is estimated to be ~ 1.5 ($Ti^{2.5+}$). Fig. 9c shows the Sr $3d_{5/2}$ binding energy calculated for the case of a bulk site as function of the valence occupation, $m$ ($2 - m$ is the charge state). The experimentally measured $3d_{5/2}$ binding energy is 133.5 eV, giving the effective ground state bulk occupancy $m_b \sim 0.1$ ($Sr^{1.9+}$). These results confirm the essentially ionic character of Sr-O bond, and the partially covalent character of Ti–O bond in STO bulk, and agree with the values found in literature [11-13]. Photoemission spectra should be more correctly interpreted in terms of the Quasi-particle (QP) spectral function including all correlation and exchange effects, whereas the LHPI model is a ground state model. However, the use of Al Kα radiation probably limits the effect of correlation and exchange.

Fig. 9b and d show, respectively, the binding energy of Ti $2p_{3/2}$ and Sr $3d_{5/2}$ calculated for Ti and Sr surface sites as a function of their valence electron occupancy $n_s$ and $m_s$. Comparison between the calculations and the experimentally measured Ti $2p_{3/2}$ and Sr $3d_{5/2}$ surface shifted binding energies, reported in Table 1 and reproduced in Fig. 9b and d, allows determination of $n_s$ and $m_s$ for the three STO surfaces investigated. In the Table 6, the effective bulk and surface electron occupancies are summarized both for Ti and Sr sites.

The uncertainty in these values is around ±0.05, which is the range of variation for the electron occupancies within the uncertainty of the binding energies (±0.1 eV as determined by the best fits to the core level spectra and shown by the shaded zones in the Fig. 9b and d). Thus, the electron occupation at surface cation sites is greater than at the bulk sites for all three surface orientations, consistent with the picture of covalency enhancement of the cation-anion bonds (Ti-O and Sr-O) in the surface plane. The relative increase in covalency of the Ti–O bond at the surface with respect to the bulk is three times bigger than that of the Sr-O bond, indicating a much stronger tendency for the O 2p orbitals to hybridize with the Ti 3d rather than with the Sr 5s. This is in qualitative agreement with the conclusions of Ghosez et al. [56] on dynamical atomic charge behaviour in the case of small ionic displacements. Furthermore, the relations $n_s^{(100)} < n_s^{(110)}$ and $m_s^{(111)} < m_s^{(100)} < m_s^{(110)}$ suggest that the surface covalency is a minimum for the (111) surface, increasing both for Ti and Sr sites, for the (100) and (110) surfaces, attaining a maximum for the latter case. The results are in good agreement with several previous theoretical works. For the (100) surface, Ellialtioglu and Wolfram [11] found an increase of the Ti 3d electron occupancy from $n_b = 1.1$ in the bulk to $n_s = 1.55$ in the surface ($\Delta n = n_s - n_b = 0.45$). For the (110) surface, Bottin et al. [57] used Density Functional Theory (DFT) calculations within the local density approximation (LDA), obtaining $\Delta n = 0.44$ for Ti sites and $\Delta m = 0.18$ for Sr sites in the SrTiO termination plane. Recently, Eglitis and Vanderblit [13] using Hartree-Fock and DFT calculations found an increase of the bonding charge of Ti–O bonds for both (100) and (110) surfaces with respect to the bulk, predicting a higher covalency in the (110) with respect to the (100) surface.

The key point in the calculations is represented by the computational methods, and the validity of the hypotheses used to determinate each term present in Eq. (1). The latter is commonly used in photoemission to represent the binding energy of a localized electron in a crystal. To calculate the free-atom ionization potential we used the Cowan program, which performs atomic multiplet calculations of electronic states in the proper crystal field symmetry using a self-consistent approach. Lindgren [58] has studied various methods for free-atom electron binding energy calculations, such as Self-Consistent-Field (SCF), all-order perturbative, Green's function and DFT methods, showing that the obtained values have a relative dispersion of only a few percent.

The Madelung energy has been obtained using the Tosi's method within the framework of the point-ion model, whose main hypothesis is the approximation of the ion's charge density by a point charge. This will be reasonable if the electron distribution of each ion is spherically symmetric, which is a good approximation in the case of Sr ions because of the s-character of the partially filled orbitals. The most general case involves the inclusion of higher order moments of the electronic charge density, and thus the Madelung energy can be obtained from a Taylor expansion. The value calculated within the framework of the point-ion model represents the first order term (monopole). Birkholz [59] demonstrated that higher order moments have to be included in the calculations only for systems where the point symmetry causes the occurrence of crystal electric field at ion's positions, and thus inducing a local net polarization. For the centro-symmetric $O_h$ structure such fields cannot occur [59] and only the sum of the point charge potentials gives a non-zero result. Thus for STO, higher order moments are forbidden from symmetry, and the value obtained within the point-ion model is a good approximation to the Madelung energy.

Table 6
Effective state of charge and effective bulk and surface electron occupancies for Ti and Sr sites (the uncertainty is around 0.05).

| | Ti | | Sr | |
|---|---|---|---|---|
| | $Q$ | $n(\Delta n)$ | $Q$ | $m(\Delta m)$ |
| Bulk | +2.5 | 1.5 | +1.9 | 0.1 |
| (111) | – | – | +1.81 | 0.19(0.09) |
| (100) | +2 | 2(0.5) | +1.74 | 0.26(0.16) |
| (110) | +1.8 | 2.2(0.7) | +1.69 | 0.31(0.21) |

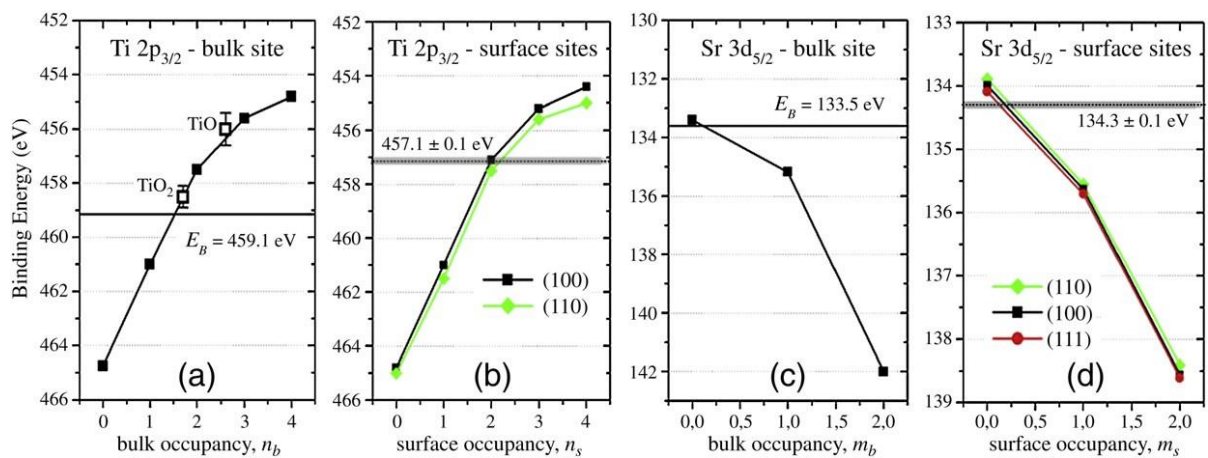

Fig. 9. Theoretical calculations of the binding energies as a function of the valence electron occupancy for: (a) Ti $2p_{3/2}$ bulk site (as comparison the experimental values for Ti $2p_{3/2}$ binding energies in TiO$_2$ and TiO are showed; the effective d-electron population for Ti ions in TiO$_2$ and TiO is taken from [8]); (b) Sr $3d_{5/2}$ bulk site; (c) Ti $2p_{3/2}$ surface sites; (d) Sr $3d_{5/2}$ surface sites.

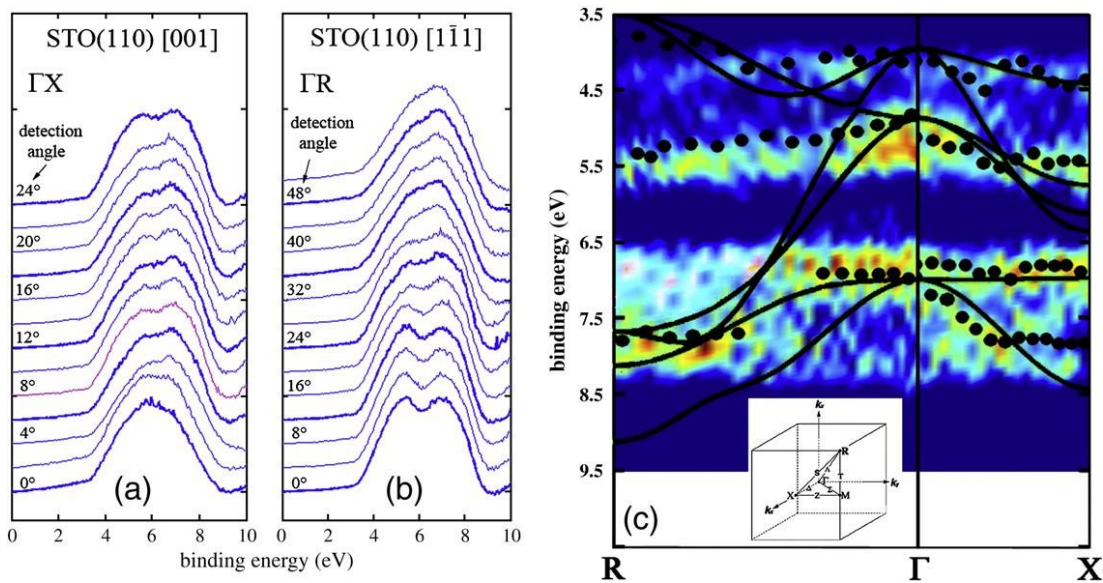

Fig. 10. (a) UPS raw spectra measured along [001] direction on a (110) surface giving the dispersion along the ΓX line; (b) UPS raw spectra measured along [1 11] direction on a (110) surface giving the dispersion along ΓR line; (c) Colour coded second derivative experimental valence band mapping along Γ-X and Γ-R lines in the bulk Brillouin zone and the same using Gaussian multipeak fits (black dots). The solid black curves represent the theoretical band structure calculated by Kotani et al. [61] using a scGW approach. Inset (c): schematic of the high symmetry directions of the Brillouin zone of a cubic lattice.

We conclude that our results and the relative discussion are sufficiently robust both with respect to the theoretical methods used for the binding energies calculation and to the theoretical framework within they are computed.

For the (100) surface, LEED patterns and surface shifts in O 1s and Sr 3d XPS peaks are consistent with a majority SrO-termination. However, the presence of a Ti 2p shifted peak (i.e. surface peak) suggests the existence of a small proportion of TiO$_2$-termination (as shown in Section 3.2). The Ti and Sr valence electron occupancy increases at the surface plane with respect to the bulk: Ti$^{2+}$ ($\Delta n = 0.5$) and Sr$^{1.74+}$ ($\Delta m = 0.16$) are the effective charge states on the (100) surface. The existence of surface rumpling should be integrated into this picture, since it would modify the bond hybridization. The (111) surface is predominantly charge compensated SrO$_3^{4-}$ terminated, as shown by the fact that only Sr 3d and O 1s peaks show a surface component. The Sr valence electron occupancy increases at the surface plane respect to the bulk: Sr$^{1.81+}$ ($\Delta m = 0.09$) is the effective charge state at the (111) surface. Thus, more covalency appears in the Sr-O bond at the surface compared to the bulk where it is more ionic (Sr$^{1.9+}$). At the (110) surface the Ti and Sr valence electron occupancy increases with respect to the bulk: Ti$^{1.8+}$ ($\Delta n = 0.7$) and Sr$^{1.69+}$ ($\Delta m = 0.21$) are the effective charge states in the (110) surface. These results confirm the validity of the SEC picture, and show that the Ti–O and Sr-O bonds on the (110) plane are characterized by higher covalency with respect to the (100) and (111) surfaces. The simple model used to quantify the surface enhanced covalency demonstrates the link between covalency and the Madelung potential.

### 4.4. Surface and bulk charge fluctuations

The measured valence band widths compare well with the UPS results of Chambers et al. [45] and their self-consistent quasi-particle calculations. As in [45], a small but significant photoelectron tail extends about 1 eV into the gap region. This is as yet unexplained. We note that Chambers et al. [45] do not observe such a broadening in XPS valence band spectra. One explanation may be due to the finite $k$ perpendicular resolution of the photoelectron final state which can significantly broaden the measured spectral function at UPS energies [60].

To check the valence band data, we have measured band structure along two high symmetry directions (ΓR and ΓX). The results are shown in Fig. 10. The principal bands agree well with the self-consistent GW calculations by Kotani et al. [61], with the exception of the supposed O 2p bands dispersing along ΓR between 4.5 and 8 eV. This may be linked to increased p-d hybridization; a greater d character in these bands may well induce strong cross-section effects [62]. The top of the VBM at the R point, combined with the expected conduction band offset at the Γ point, given the Nb doping level, return an energy gap of 3.25 ± 0.10 eV, in excellent agreement with optical absorption measurements for the indirect band gap.

In light of the measured VB data, we can discuss the charge fluctuation trends predicted by Bocquet et al. [8], in particular to determine how the charge fluctuation mechanism evolves going from the bulk to the surface. Bocquet and co-workers classified the STO bulk in the charge transfer regime (Δ=4.0 eVb $U$ = 4.5 eV), and noted that Δ rapidly increases (+2 eV for each added 1 d-electron), while $U$ slowly decreases (−0.5 eV for each added 1 d-electron), as a function of the electron occupancy of Ti 3d states. We found that the increase of the electron occupancy, Δn, is 0.5 for the (100) surface and 0.7 for the (110) surface. In both cases, Δ becomes greater than $U$ and thus the two surfaces tend more towards the Mott-Hubbard regime. Therefore, while into the bulk the metal d-ligand p charge transfers are energetically favourable, on the (100) and (110) surfaces charge fluctuations between two neighbouring Ti ions of the type $d^n d^n_j \leftrightarrow d^{n-1}_i d^{n+1}$ become more important. Indeed the higher covalent character will stiffen the bonding making ligand-cation charge transfer more difficult.

5. Conclusion

We have investigated the bulk and surface covalency of Sr–O and Ti–O bonds in SrTiO$_3$(100), (110) and (111). We provide quantitative confirmation of the essentially ionic character of Sr–O bond, and the partially covalent character of Ti–O bond in STO bulk: the effective ground state bulk occupancy for Ti ions has been evaluated to be $n_b \sim 1.5$ (Ti$^{2.5+}$), and for Sr ions to be $m_b \sim 0.1$ (Sr$^{1.9+}$). The electron occupation at surface cation sites is evaluated for all the three surfaces to be greater than that at bulk sites. This increase is consistent with the picture of an enhancement of the covalency of cation-anion bonds (Ti–O and Sr–O) at the surface. The relations $n_s^{(100)} = 2 b n_s^{(110)} = 2.2$ and $m_s^{(111)} = 0.19 b m_s^{(100)} = 0.26 b m_s^{(110)} = 0.31$ suggest that the surface covalency is a minimum for the (111) surface, and increases for both Ti–O and Sr–O bonds, for the (100) and (110) surfaces, attaining a maximum in the latter case. Bond covalency is also studied in relation to measured valence band widths. Using the variations of Δ and $U$ as a function of the Ti 3d states occupancy we suggest that on the (100) and (110) surfaces charge fluctuations between two neighbouring Ti ions are more important, while in the bulk the metal $d$-ligand $p$ charge transfers are more energetically favourable. Our results also underline that correlations between small surface structural changes (relaxation or rumpling) and covalency should be more systematically measured by experiment, providing valuable comparison with first principles calculations such as those presented in [13]. Quasi-particle spectral function appears necessary in order to fully understand the valence band spectra and some careful investigation of the reasons behind the smearing of the valence band structure at low photon energies should be done.

Acknowledgements

The authors would like to thank O. Renault, J. Leroy and B. Delomez for technical assistance, R. Gusmeroli for the precious aid with the Cowan's program, N. Vast and A. Tagliaferri for fruitful discussions. This work was financially supported by the European commission under contract Nr. NMP3-CT-2005-013862 (INCEMS) and by the French National Research Agency (ANR) through the "Recherche Technologique de Base" Program.